\documentclass[prd,twocolumn,aps,nofootinbib,showpacs]{revtex4}
\usepackage{amsmath}
\usepackage{amssymb}
\usepackage{graphicx}
\usepackage{color}
\usepackage{bm}
\usepackage{times}
\usepackage{hyperref}
\hypersetup{
  colorlinks=true,
  citecolor=blue,
  linkcolor=blue,
  urlcolor=blue}

\usepackage{epsfig}
\usepackage{dcolumn}
\usepackage{float}

\usepackage{epsfig}
\usepackage{dcolumn}
\usepackage{morefloats}
\def\be{\begin{equation}}
\def\ee{\end{equation}}
\def\bea{\begin{eqnarray}}
\def\eea{\end{eqnarray}}
\def\gsim{\ \rlap{\raise 2pt\hbox{$>$}}{\lower 2pt \hbox{$\sim$}}\ }
\def\lsim{\ \rlap{\raise 2pt\hbox{$<$}}{\lower 2pt \hbox{$\sim$}}\ }
\def\dslash{\kern-4pt \not{\hbox{\kern-2pt $\partial$}}}
\def\pslash{\not{\hbox{\kern-2pt p}}}

\newcommand{\dcp}{\delta_{CP}}
\newcommand{\nova}{NO$\nu$A\ }

\begin{document}

\renewcommand{\arraystretch}{2}
\DeclareGraphicsExtensions{.eps,.ps}


\title{Implications of the latest NO$ \nu $A results}


\author{Srubabati Goswami}
\email[Email Address: ]{sruba@prl.res.in}
\affiliation{
Physical Research Laboratory, Navrangpura,
Ahmedabad 380 009, India}

\author{Newton Nath}
\email[Email Address: ]{newton@prl.res.in}
\affiliation{
Physical Research Laboratory, Navrangpura,
Ahmedabad 380 009, India}
\affiliation{Indian Institute of Technology, Gandhinagar, Ahmedabad--382424, India}

\begin{abstract}
We discuss the implications of the latest  \nova results 
on the measurement of  $\nu_\mu - \nu_e$ conversions. 
From a combined analysis of the disappearance and appearance data, 
the preferred solutions reported by \nova are normal hierarchy (NH) 
with two degenerate best-fit points
one in the lower octant (LO)  and $\dcp = 1.48\pi$ whereas  the other in 
the higher octant (HO) with $\dcp = 0.74\pi$. 
There is also another solution with inverted hierarchy (IH) 
which is $0.46 \sigma$ away from the best-fit. 
We discuss and quantify the possibility of resolving these degeneracies 
by inclusion of \nova antineutrino runs. 
We show that if the true solution corresponds to (NH, LO, $\sim 1.48\pi$) then 
future data from \nova comprising of 3 years of neutrino and 3 years 
of antineutrino run 
will be able to resolve the degenerate solutions 
at 95.45\% C.L.
However, if the true solution corresponds to (NH, HO, $\sim 0.74\pi$), 
a wrong hierarchy-wrong $\dcp$ solution remains unresolved even at 
68\% C.L. with the full $3\nu+3\bar{\nu}$ projected run of \nova. 
Same is the case if the IH solution turns out to be the true solution.  
We further show that DUNE (10 kton)  will be able to resolve these 
degenerate solutions at 
99.73\% C.L. with only 1 year of neutrino and one year of  
antineutrino 
run [1+1] while 5+5 years of DUNE data can resolve the degeneracies at 
99.99\% C.L.

\end{abstract}
\maketitle
\section{Introduction}

Neutrino  oscillation physics 
is standing at a very interesting  juncture, 
awaiting the determination  of the three remaining unknown parameters
-- the neutrino mass hierarchy, octant of the 2-3 mixing angle and  
the CP phase $\dcp$. Neutrino oscillation implies that the flavour 
states ($\nu_e,\nu_\mu,\nu_\tau$) of the neutrinos are not the same
as the mass states ($\nu_1,\nu_2,\nu_3$).   
Depending on the relative ordering of 
the third mass eigenstate there can be two 
possible mass orderings or hierarchies. If 
$m_3 > m_2 > m_1$ we call it normal hierarchy (NH) whereas  
$m_3 < m_1 \approx m_2$ is termed as inverted hierarchy (IH). 
If the mixing angle $\theta_{23}$ is not exactly $\pi/4$ then 
there can be two options: 
$\theta_{23} < \pi/4$ known as  the  lower
octant (LO) or it is  $>\pi/4$ called the higher octant (HO). 
For the CP phase $\dcp$ the best-fit comes close to $270^\circ$ 
while  at $3\sigma$ the whole range from 
$0-2\pi$ remains allowed from global oscillation analysis  
\cite{Capozzi:2016rtj,Esteban:2016qun}. These analyses have 
not yet included the recent \nova results reported in 
\cite{Adamson:2017gxd}. 
The other oscillation parameters, namely, 
the two mass squared differences ($\Delta m^2_{21} = m_2^2 -m_1^2$, $|\Delta m^2_{31}|=|m_3^2 - m_1^2| $,
the two leptonic mixing angles
($\theta_{12}$ $\theta_{13}$) have been determined 
with a good  precision by the global analysis of data from
oscillation experiments \cite{Capozzi:2016rtj,Esteban:2016qun}. 
At present the focus is on the  currently running   
beam based, off-axis, experiments T2K and \nova  which have declared their 
initial results on measurement of $\dcp$  
\cite{Abe:2013hdq,Adamson:2016tbq,Abe:2017uxa,Adamson:2017gxd}.
These experiments, 
in conjunction with the reactor measurement of the
1-3 mixing angle $\theta_{13}$
\cite{dayabay_t13,reno_t13,dchooz_1406}, are expected to shed light on the 
above unknown parameters and provide  the direction for future experiments. 

The major problem, in the accurate determination 
of the above unknowns,  in long-baseline experiments
is the parameter degeneracies 
\cite{barger,lisi,intrinsic,degeneracy1,Minakata:2001qm,degeneracy4,minakata}. 
For recent studies see e.g. \cite{ suprabh_t2knova,usoctant,suprabhlbnelbno,
Ghosh:2013zna,Prakash:2013dua,Agarwalla:2014fva,
Ghosh:2014dba,Ghosh:2014rna,C.:2014ika,Deepthi:2014iya,Coloma,Ghosh:2015tan,Goswami:2016auo,
Soumya:2016aif}.
Recently,  it was shown that if 
we classify the solutions in terms of the three unknown parameters 
hierarchy, octant and $\dcp$ then 
there can   be a total of
eight degenerate solutions 
involving wrong hierarchy and/or wrong octant and/or wrong $\dcp$
\cite{Ghosh:2015ena} 
\footnote{Note that these solutions are somewhat different 
from the original eightfold degeneracy discussed in \cite{barger}. 
Specifically the  intrinsic $\theta_{13} -\dcp$ 
degeneracy \cite{intrinsic} which is a part of the eightfold 
degeneracy 
is now resolved.}.    
This  can be best studied in terms of a 
generalized hierarchy-$\theta_{23}- \dcp$ 
degeneracy, presented as contours in $\theta_{23} - \dcp$ plane 
\cite{Ghosh:2015ena}. 
Which of these degeneracies actually exists depends on the baseline 
of the experiment and the true parameter values.   

The most recent \nova analysis,
incorporating both appearance and disappearance 
channel data in neutrino mode,
shows that there are two best-fit points occurring 
for normal hierarchy \cite{Adamson:2017gxd}. 
These are (i)  $\sin^2\theta_{23} = 0.404, \dcp = 1.48\pi = 266.4^\circ$ 
and  (ii) $\sin^2\theta_{23} = 0.623, \dcp = 0.74\pi = 133.2^\circ$. 
The two solutions correspond to  the degeneracy in neutrino 
probability between opposite octants due to different values of 
$\dcp$. One of them could be the true solution whereas the other 
corresponds to a degenerate solution with wrong-octant and wrong-$\dcp$. 
\nova also reports  a solution with IH, HO and $\dcp = 270^\circ$   
which
is only $0.46\sigma$ away from the best-fit. 
This corresponds to the wrong-hierarchy solution with wrong 
octant and right $\dcp$ for (i) and wrong-hierarchy right-octant and wrong
$\dcp$ 
solution with respect 
to (ii). Inverted hierarchy with $\theta_{23}$ belonging to lower 
octant is disfavoured at greater than 90\% C.L. irrespective of the 
value of $\dcp$.
The T2K results on the measurement of $\dcp$ on the other hand gave  
a hint of $\dcp \sim 270^\circ$ from neutrino data \cite{Abe:2013hdq}.
Recently T2K has published their results for search of CP violation 
using both neutrino and antineutrino runs and including appearance 
and disappearance channel data \cite{Abe:2017uxa}. 
The best-fit obtained is for 
NH at $\dcp = 1.43\pi = 258^\circ$.   
It has been argued in \cite{Ghosh:2015ena}, 
based on the degeneracies in the probabilities and a simulation
of T2K data with its projected full power,
that a hint of $\dcp  \sim 270^\circ$ in
neutrino channel will 
imply the hierarchy to be NH and octant to be HO while such 
an indication for the  antineutrino channel will imply the octant to be LO. 
Thus an unambiguous hint for $\dcp = 270^\circ$ in both neutrino and 
antineutrino channel will signify the mixing angle $\theta_{23}$ to
be maximal. 
T2K has indicated that $\dcp \sim 270^\circ$ in both
neutrino and antineutrino 
mode and the $\theta_{23}$ from analysis of T2K data indeed
comes out to 
be $45^\circ$ 
\footnote{T2K has recently published separate fits to
the neutrino and the  antineutrino data in 
the muon disappearance channel and while the neutrino best-fit continues to be $\sin^2\theta_{23} = 0.51$ 
for the case of antineutrino fit the best fit value is 0.42. 
However, at 68\% C.L. the $\sin^2\theta_{23}$ range 
from the neutrino fit is contained in that obtained from
the antineutrino fit \cite{Abe:2017uxa}.}.  
Thus there is some mismatch between the  T2K and \nova results 
and more data may be able to  resolve these issues. 

In this paper, we study the implications of the  latest \nova results and 
to what extent the different degenerate solutions can be resolved 
by future runs of \nova. The projected runtime of \nova is 3 years in
neutrinos and 3 years in antineutrinos. 
After declaring the results with one year neutrino run 
\nova is currently running in the antineutrino mode. 
It is well studied that the addition of antineutrino 
runs can resolve the wrong octant solutions
\cite{suprabhoctant,minakata_cp,Ghosh:2015ena,Nath:2015kjg} and  
can give rise to an improved precision in $\dcp$. 
In view of this  we discuss if it is profitable for \nova to continue with the 
antineutrino runs for three years or they should switch over to 
neutrino run after one year of antineutrino run. 
We also investigate how  the next generation 
experiment DUNE at Fermilab can further improve the 
prospect of  resolution of these 
degeneracies. 

The plan of the paper is as follows -- 
in section (\ref{sec:expt}) we present the experimental specifications that 
have been used in our numerical simulation. 
In the next section (\ref{sec:nova_prob_deg}), we  demonstrate the probabilities for the 
\nova and the DUNE baselines
and show the occurrence of degeneracies at the probability level. 
We also show the fate of the degenerate solutions for different 
run times of \nova assuming 
each of the three solutions -- NH-LO, NH-HO and IH-HO as the 
true solutions, in section (\ref{sec:nova}). Finally, in section (\ref{sec:nova_dune}) we present the results combining DUNE with \nova and estimate at what significance the degeneracies can be removed. We summarize our findings in section (\ref{sec:conclusion}).

\vspace{-5mm}
\section{ Specifications of the experiment}\label{sec:expt}
\vspace{-3mm}
In the numerical simulations of the \nova\ (NuMI Off-Axis $ \nu_{e} $-Appearance, Fermilab) and DUNE (Deep Underground Neutrino Experiment) data, we use the GLoBES package~\cite{globes1,globes2} along with the required auxiliary files~\cite{messier_xsec,paschos_xsec}. 
The \nova\ experiment sends muon neutrino beams through two detectors-- one the near detector (at Fermilab ) and the far detector (northern Minnesota). The far detector is placed  810 km away from the source by making an off-axis angle of $0.8^\circ  $. The detector is  a totally active scintillator detector (TASD) with a volume of 14 kton. 
 \nova\ has planned to reach 700 kW beam power which 
corresponds to $ 6 \times 10^{20} $ POT (proton on target) per year. The current reach is 
560 kW.  The experiment will run for 
$ (3+ 3)$ years $ (\nu+\overline{\nu}) $. 
In our simulation of \nova\ data, we consider the 
reoptimized \nova\ set up from Refs.\cite{sanjib_glade,Kyoto2012nova}.
For the numerical analysis of DUNE data, we consider 10 kton
far detector mass which will be 1300 km away on-axis
from the neutrino source 
and is planned to be placed at the Sanford Lab in South Dakota. 
The source of this project is also based on the NuMI beam at Fermilab. The 
beam flux  will be peaked at 2.5 GeV. 
The current plan of DUNE collaboration is to have 
an initial beam of 1.2 MW which will be upgraded to 2.3 MW \cite{Acciarri:2015uup}. In this analysis, we use neutrino flux from \cite{dancherdack} corresponding to 1.2 MW beam power and 120 GeV proton energy which will produce $ 10^{21} $ POT per year .
Systematic errors are taken into account using the method of pulls~\cite{pulls_gg,pull_lisi} as outlined in Ref.\cite{ushier}. We have added a  $5\%$ prior on $ \sin^2 2\theta_{13}$. 

The values of  the oscillation 
parameters used in our numerical simulation are given
in tab.~\ref{param_values}.
 \begin{table}
 \begin{tabular}{|c|c|c|}
 \hline
 Osc. param. & True Values &   Test Values \\
 \hline   
$ \sin^{2}2 \theta_{13} $ & 0.085 & 0.075 -- 0.095 \\
$ \sin^{2}\theta_{12} $ & 0.306 & Fixed \\
$ \sin^{2}\theta_{23} $ &  LO=0.40, HO=0.62  & 0.22 --0.70 \\
$\Delta_{21} ({\rm eV^{2}}) $ &  7.50 $ \times 10^{-5} $  &  Fixed \\
$\Delta m^2_{31}({\rm eV^{2}}) $ & 2.40 $ \times 10^{-3}  $ & (2.30 -- 2.70) $ \times 10^{-3} $ \\
$ \delta_{CP} $ &  $ 135^\circ,~ 270^\circ $  & $0^\circ $ -- $ 360^\circ $ \\  
%
  \hline
 \end{tabular}
 \caption{Oscillation parameter values that are considered in 
the numerical analysis  
\cite{Capozzi:2016rtj,Esteban:2016qun}. The $\sin^2\theta_{23}$ and 
$\dcp$ values are motivated from the latest \nova results 
\cite{Adamson:2017gxd}.  
}

 \label{param_values}
 \end{table}
\section{Probability level discussion}\label{sec:nova_prob_deg}
\begin{figure*}[htbp]
        \begin{tabular}{lr}
\includegraphics[height=5cm,width=7cm]{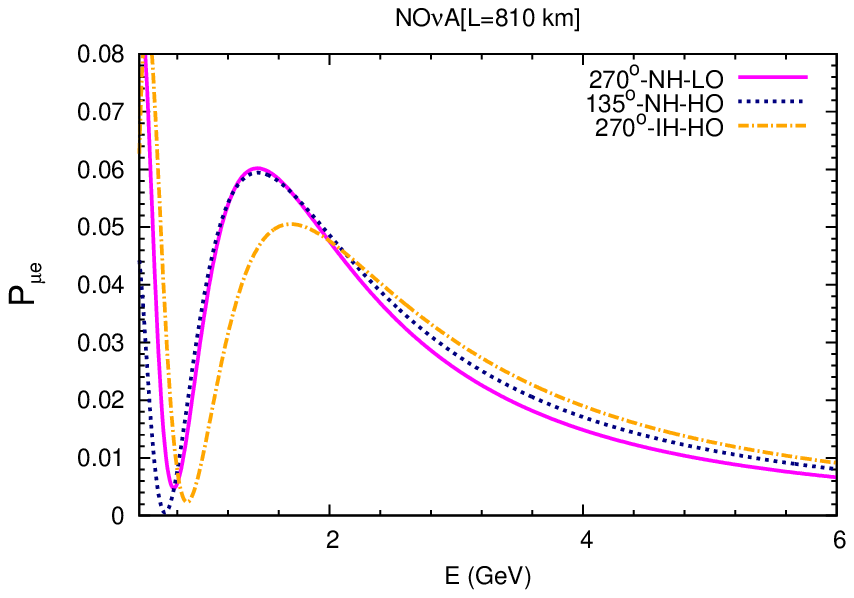} 
\includegraphics[height=5cm,width=7cm]{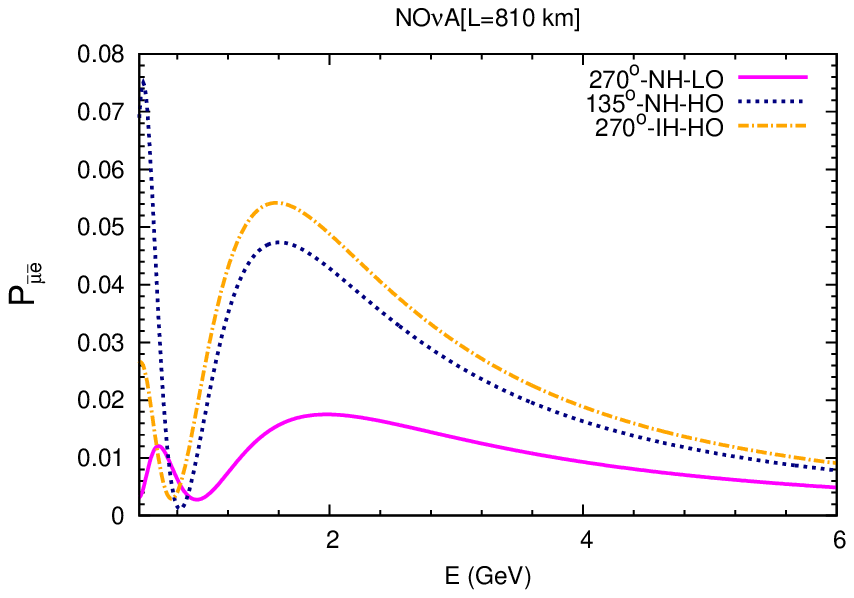} \\
\includegraphics[height=5cm,width=7cm]{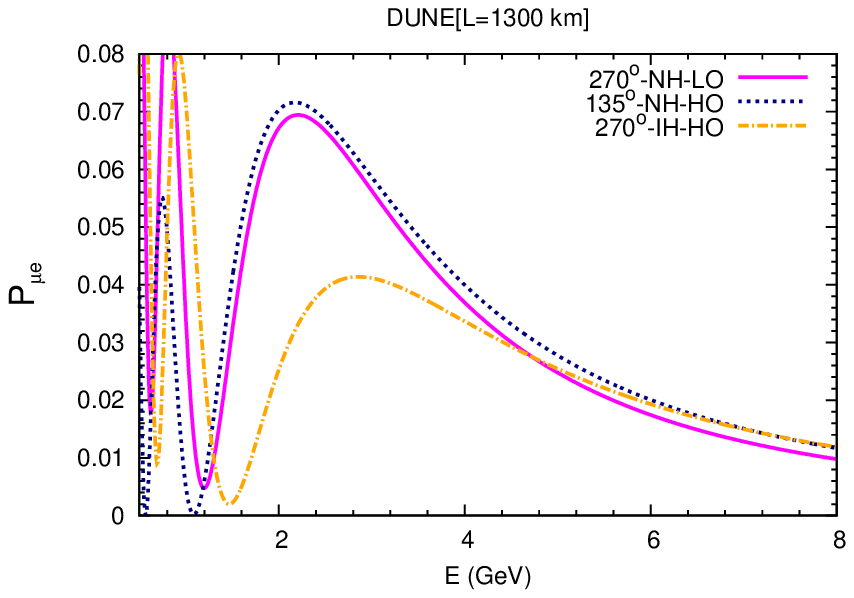} 
\includegraphics[height=5cm,width=7cm]{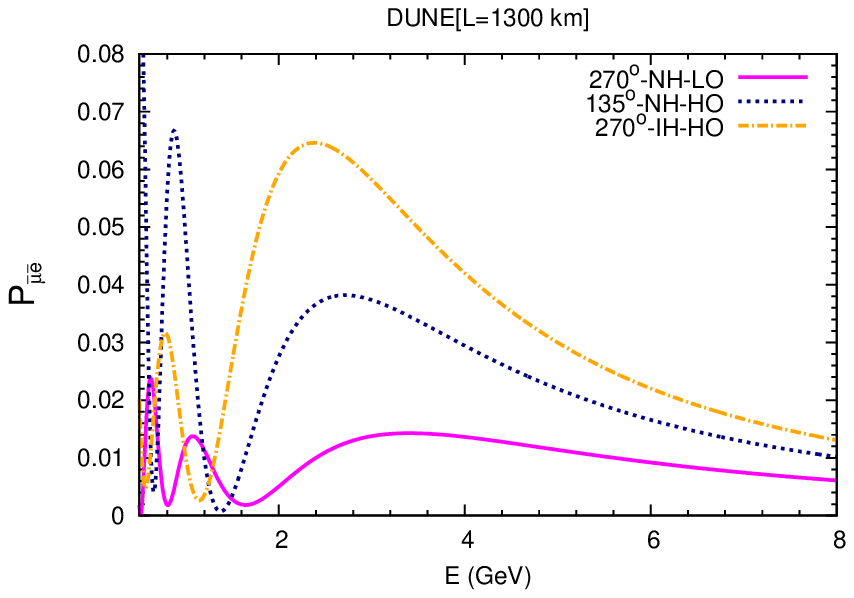}      
        \end{tabular}
\vspace{-0.35cm}
\caption{\footnotesize The oscillation probability  $P_{\mu e}$ 
as a function of energy. Here, we consider $ \sin^{2}\theta_{23} = 0.4(0.62) $ for LO (HO). The top (bottom) panel is for NO$ \nu $A (DUNE).
The left panel is for neutrinos, while the right panel is for antineutrinos.
 }
\label{fig:novaprob} 
\end{figure*}
The relevant expressions for the 
oscillation  and survival probabilities in matter of constant density
(a good approximation for the baselines concerned) are 
as follows \cite{akhmedov,cervera,freund}: 
\begin{align} 
 P_{\mu e} & =  \alpha^2 \sin^22\theta_{12} c_{23}^2 \frac{\sin^2\hat{A}\Delta}{\hat{A}^2}          + 4 s_{13}^2 s_{23}^2 \frac{\sin^2(\hat{A} - 1) \Delta}{(\hat{A} - 1)^2}   \nonumber \\
& + 2 \alpha s_{13} \sin2\theta_{12} \sin2\theta_{23}  \nonumber \\
& 
\times \cos(\Delta +  \delta_{CP}) 
\frac{\sin\hat{A}\Delta}{\hat{A}} \frac{\sin(\hat{A} - 1)\Delta}{\hat{A} - 1} \label{eq:matter_prob_app_ch4} \\ 
P_{\mu\mu} & = 1 - \sin^2 2\theta_{23} \, \sin^2\Delta + \mathcal{O}(\alpha, s_{13}) \label{eq:matter_prob_disapp_ch4}
\end{align}
where, $ s_{ij}(c_{ij})=\sin \theta_{ij}(\cos \theta_{ij}) $ for 
$ j>i $ ($ i,j = 1,2,3 $), 
$ \Delta =  \Delta m^{2}_{31} L / 4 E $ , 
and $ \hat{A} = A/ \Delta m^{2}_{31} $, 
with $A= 2EV=0.76 \times 10^{-4} \rho (\dfrac{g}{cc}) \times E(GeV)$, 
being the Wolfenstein matter term. The antineutrino oscillation probability can be obtained by replacing 
$\delta_{CP} \rightarrow - \delta_{CP}  $ and $ V \rightarrow  - V $. 
As we have discussed earlier, $ \Delta m^{2}_{31} > 0 $ for NH and $ \Delta m^{2}_{31} < 0 $  for IH. The matter term $ A $ is positive for  the neutrinos
and negative for the antineutrinos. Therefore, in the case of  the 
neutrino, $ \hat{A} > 0 $ for NH and  $ \hat{A} < 0 $ for IH whereas reverse is true for the antineutrinos.

In figure~\ref{fig:novaprob} we discuss the oscillation probability 
$P_{\mu e}$ vs energy for the three different sets of hierarchy, octant and 
$\dcp$. The top (bottom) panels are for NO$\nu$A (DUNE) baseline whereas
the  left (right) panel represents the neutrino (antineutrino) probability. 
The pink-solid, blue-dotted and orange-dash dotted curves are for 
270$ ^{\circ} $-NH-LO, 135$ ^{\circ} $-NH-HO 
and 270$ ^{\circ} $-IH-HO respectively.
The values of $\sin^2\theta_{23}$ for LO and HO correspond to 0.40 and 0.62
respectively. We consider these values because the current combined analysis of
NO$ \nu $A $ \nu_{e} $-appearance and $ \nu_{\mu} $-disappearance data
favors these values of  $\theta_{23}$ and 
$\dcp$ close to  these points \cite{Adamson:2017gxd}.
We notice from the left panel that for the neutrinos, 
at energy around E $\sim$2 GeV, for which the flux  
peaks for  NO$ \nu $A, all the three curves exhibit a degeneracy.
However, whereas the two NH probabilities are almost the 
same over the whole energy range, the IH probability differs from 
these, specially for energies $<$ 2 GeV. 
On the other hand, as can be seen from the right panel, 
for the antineutrinos,  the LO curve is well separated from the two HO curves 
over most of the energy range. This signifies that if the true point is 270$ ^{\circ} $-NH-LO then addition of antineutrino  information can differentiate between the 
degenerates points with opposite octants clearly.  

In the bottom panels, we present the probabilities 
for DUNE.  For this case, due to enhanced matter effect the 
neutrino probabilities are higher, but still  
there exists degeneracy between 270$ ^{\circ} $-NH-LO and 135$ ^{\circ} $-NH-HO 
whereas 270$ ^{\circ} $-IH-HO  differs from these over 
a larger energy range. However, for the antineutrinos, all the three cases are
seen to be well separated. Thus, addition of DUNE antineutrino run is expected 
to resolve  the wrong octant and wrong hierarchy solutions efficiently.    

\section{Results for \nova} \label{sec:nova}
\begin{figure*}[htbp]
\vspace{-1.0cm}
 \begin{tabular}{lr}
 \hspace*{0.3in} 
 \includegraphics[height=4.5cm,width=5.5cm]{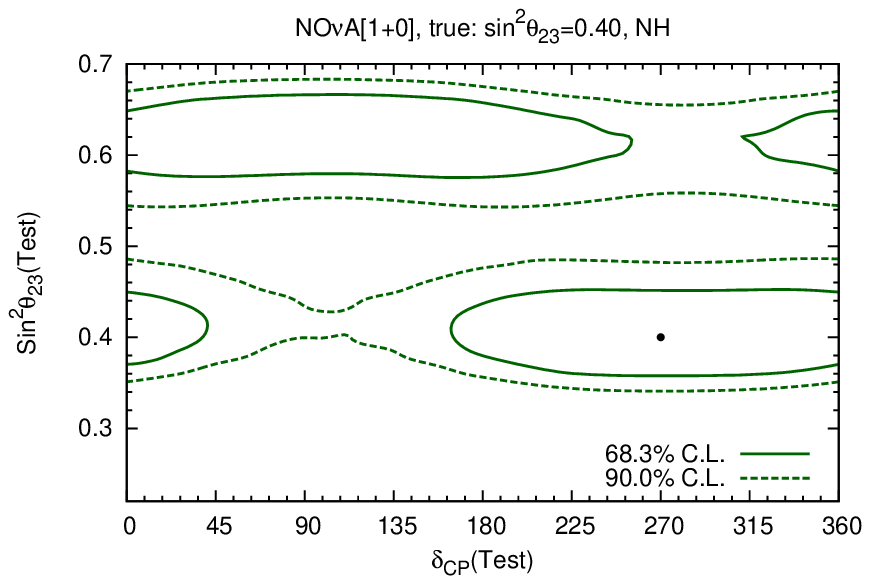} 
   \hspace*{-0.1in} 
 \includegraphics[height=4.5cm,width=5.5cm]{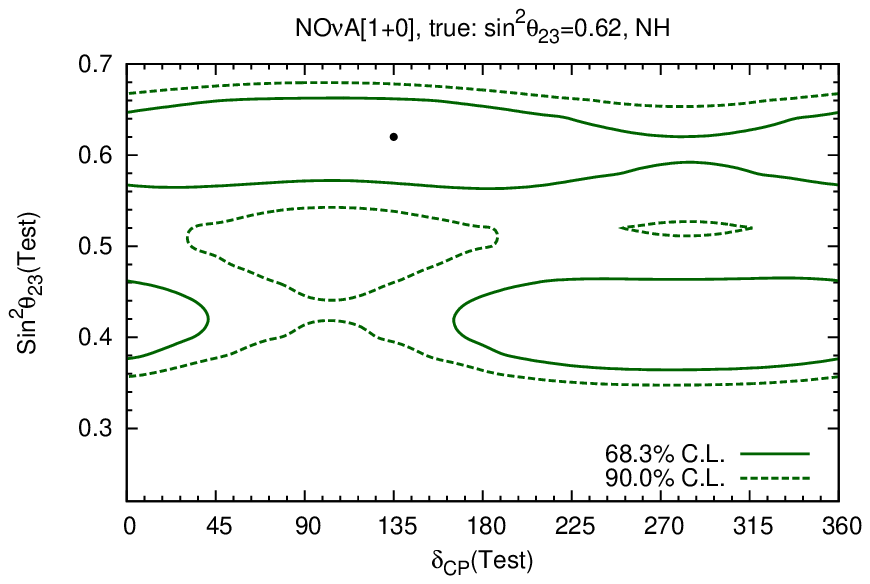}
\hspace*{-0.1in} 
\includegraphics[height=4.5cm,width=5.5cm]{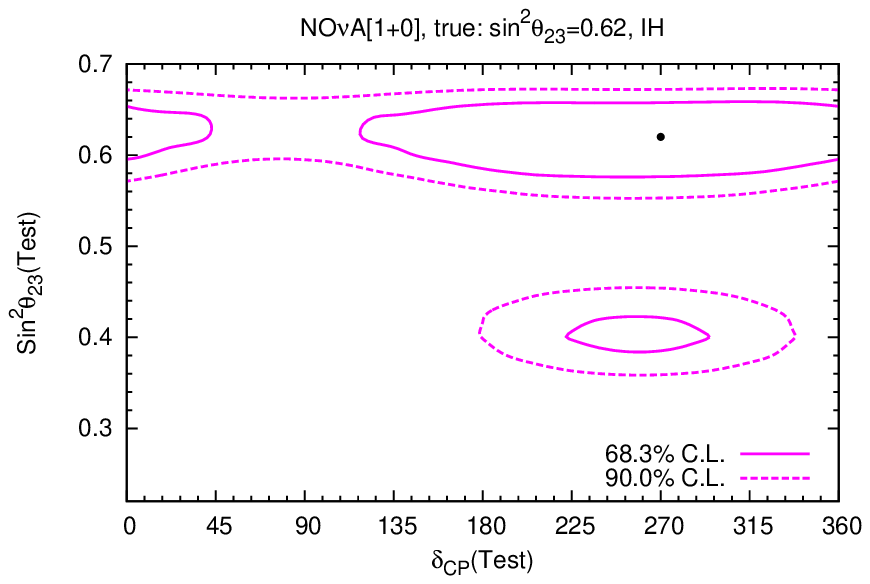} \\
 \hspace*{0.3in} 
\includegraphics[height=4.5cm,width=5.5cm]{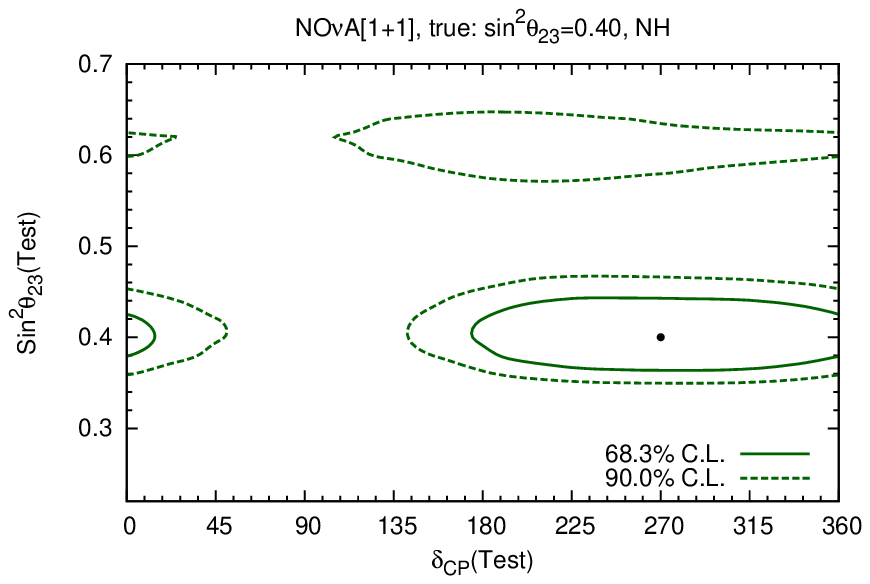} 
\hspace*{-0.1in} 
\includegraphics[height=4.5cm,width=5.5cm]{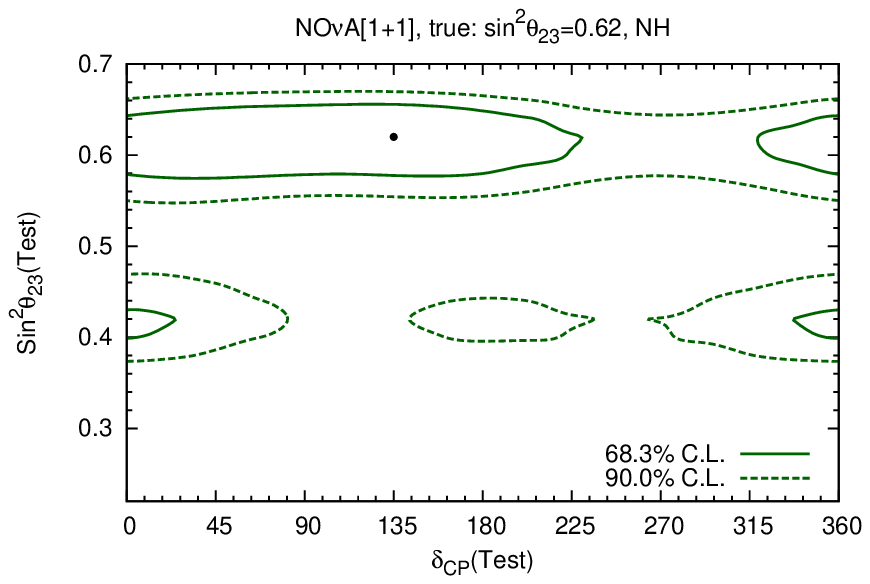}
\hspace*{-0.1in} 
\includegraphics[height=4.5cm,width=5.5cm]{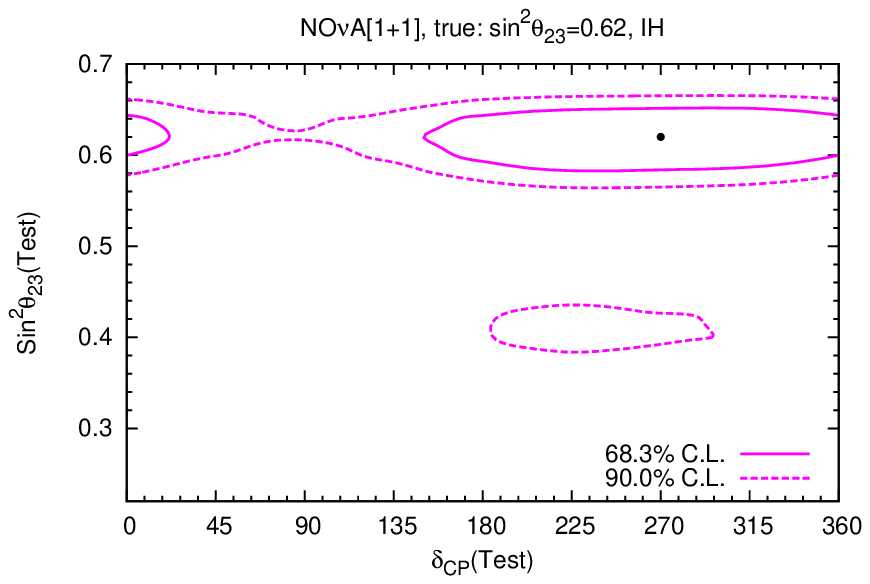}
\end{tabular}
\caption{Contour plots in test $ \sin^{2} \theta_{23} - \delta_{CP} $ plane.
True hierarchy is NH for the first and second column while it is IH for the 
third column. The black dot denotes the true value. 
}
\label{fig:nova_1_0_n_1_1}
\end{figure*}

In this section we present the allowed areas in the test 
$\sin^2\theta_{23} - \dcp$ plane from simulation of \nova 
data with different run times and identify the wrong
solutions \footnote{
In this section and throughout the paper 
we have used the following abbreviations  --- 
WH(RH) $\equiv$ wrong(right) hierarchy , 
WO(RO) $\equiv$ wrong(right) octant 
W$\dcp$(R$\dcp$) $\equiv$ wrong(right) $\dcp$. 
}

In figure \ref{fig:nova_1_0_n_1_1}, in the first row we 
present 
the allowed areas with \nova[1+0] which corresponds to running in 
neutrino mode for one year with a total pot of $6 \times 10^{20}$. 
The first plot is assuming the  true values  
to be $\dcp = 270^\circ, \sin^2\theta_{23} =0.4$ 
and NH as  both  true and test hierarchy. 
In this case, the WO solution is present  even at 68\% C.L. 
However, the  maximal mixing solution is seen to be 
disfavoured at 90\% C.L. It  also shows that at 90\% C.L. the full range of 
$\dcp$ remains  allowed. The second plot in the first row is for the true value 
$\dcp = 135^\circ, \sin^2\theta_{23} =0.62$ around the best-fit point reported 
by \nova \cite{Adamson:2017gxd} and 
for NH. This  plot also shows the presence of wrong octant 
solutions even at 
68\% C.L. The $\theta_{23}$ precision is also worse as compared to the 
earlier case and even maximal mixing is seen to be allowed for some 
of the values of $\dcp$.  The CP precision is also worse and 
even at 68\% C.L. the full range of $\dcp$ remains 
allowed  for the true solution. 
In the third plot, we consider true and test hierarchy as IH and 
the true values as $\dcp = 270^\circ, \sin^2\theta_{23} = 0.62$  
for which \nova 
reported a  local minima \cite{Adamson:2017gxd}. 
We consider this case as well in our study
since it's within $1\sigma$ of the best-fit solutions. 
We see that in this case   the LO solutions with $0<\dcp<180^\circ$ gets
disfavoured at 90\% C.L. 
Note that to obtain these plots we have not fitted the \nova data, but assumed
that the parameters for which \nova gets global or local minimum
are the 
true values and have performed our analysis with the simulated data
of one year of neutrino run. 

In the bottom row of figure \ref{fig:nova_1_0_n_1_1} we show the 
allowed areas corresponding to  1 year neutrino and 1 year antineutrino 
run of \nova. For this analysis also we take the true values as the 
best-fit points obtained by \nova as above. 
The first plot shows that the WO solution is  
resolved at 68\% C.L. with addition of one year of antineutrino data. 
At 90\% C.L. the WO solutions with CP values near $90^\circ$ 
are largely resolved but wrong octant solutions close to
$\dcp= 270^\circ$ 
remain allowed. 
The range of $\dcp$ for the true solution also gets restricted. 
The wrong octant solution is also seen to be almost  resolved 
at 68\% C.L. for the middle plot in the 2nd row.  
But at 90\% C.L. WO solutions can still be there. 
However, the maximal mixing solution gets disfavoured.  
For the true solution, at 90\% C.L., the full range of $\dcp$ 
remains allowed. But the  range of $\dcp$ in the wrong octant gets more restricted.   
The third plot is for IH and for this case also the wrong octant  
solution gets resolved at 68\% C.L. and only a small region with
WO-R$ \delta_{CP}$ remains allowed at 90\% C.L. The range of both $\theta_{23}$ and $\dcp$ gets more constricted around the true value, though the full range $0-360^\circ$
still remains allowed for $\dcp$. 

We next ask the question, given the scenario described above
after one year of neutrino run 
and one year of antineutrino run, should \nova continue running in the 
antineutrino mode or switch to neutrino mode again.


\begin{figure*}[htbp]
\vspace{-1.0cm}
 \begin{tabular}{lr}
 \hspace*{0.3in} 
 \includegraphics[height=4.5cm,width=5.5cm]{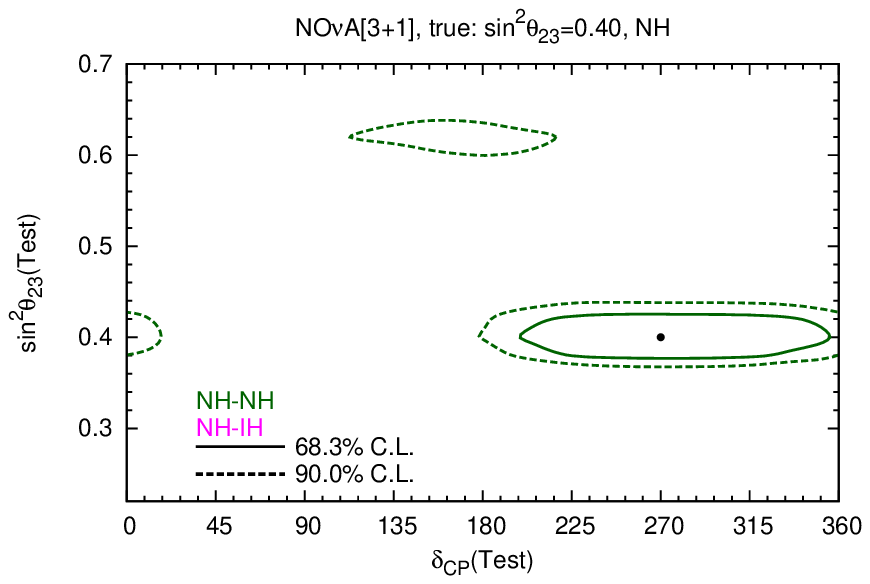} 
   \hspace*{-0.1in} 
  \includegraphics[height=4.5cm,width=5.5cm]{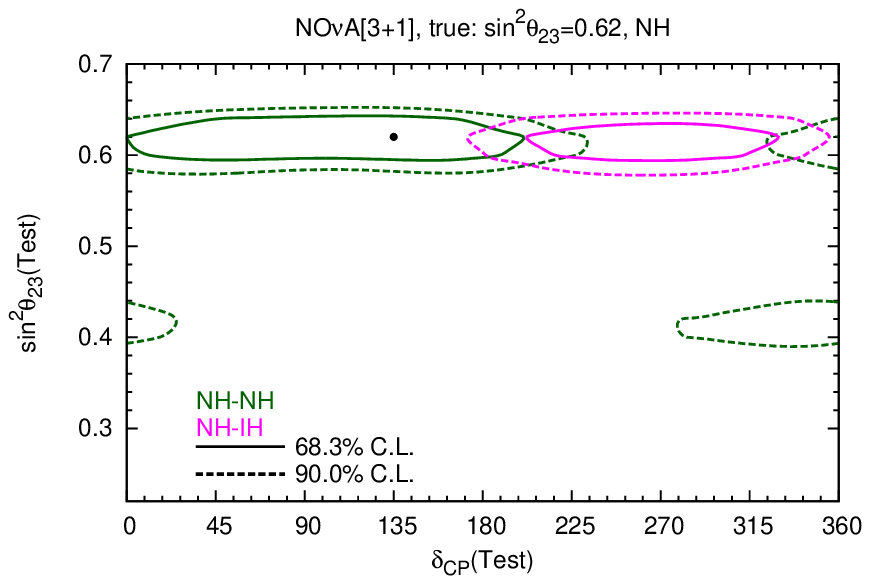} 
   \hspace*{-0.1in} 
 \includegraphics[height=4.5cm,width=5.5cm]{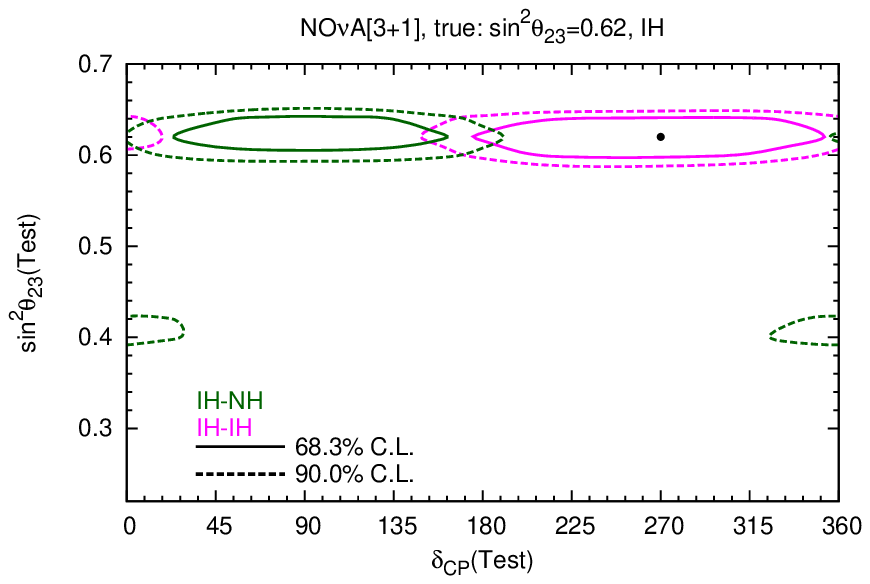} \\ 
   \hspace*{0.3in} 
  \includegraphics[height=4.5cm,width=5.5cm]{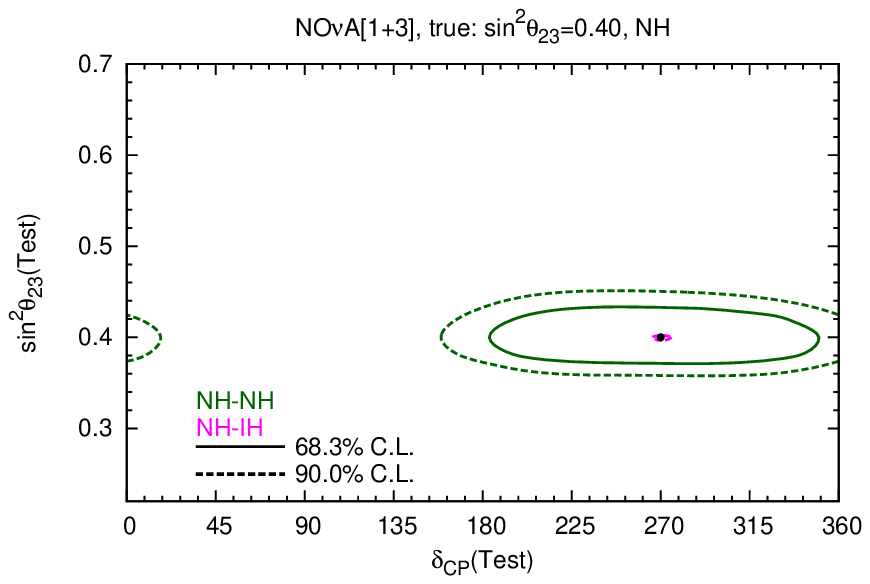} 
   \hspace*{-0.1in} 
 \includegraphics[height=4.5cm,width=5.5cm]{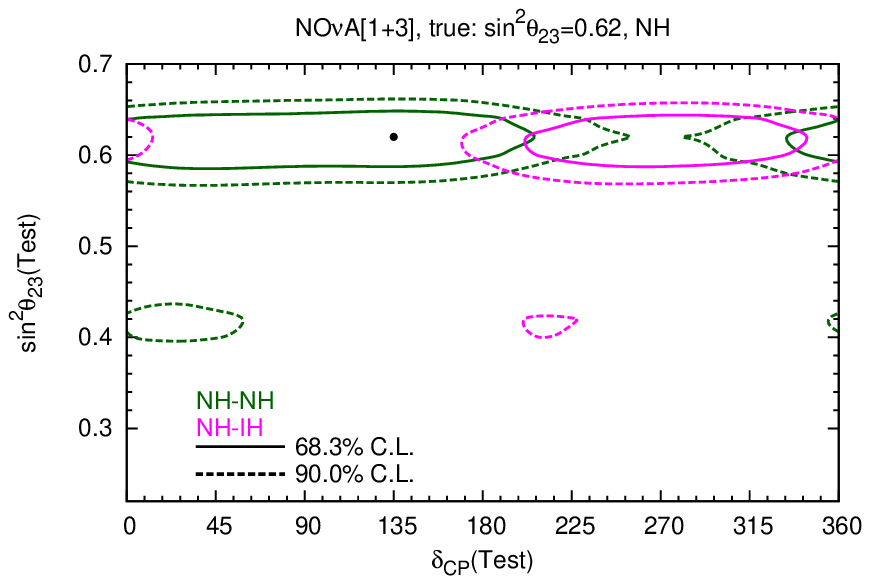} 
   \hspace*{-0.1in} 
  \includegraphics[height=4.5cm,width=5.5cm]{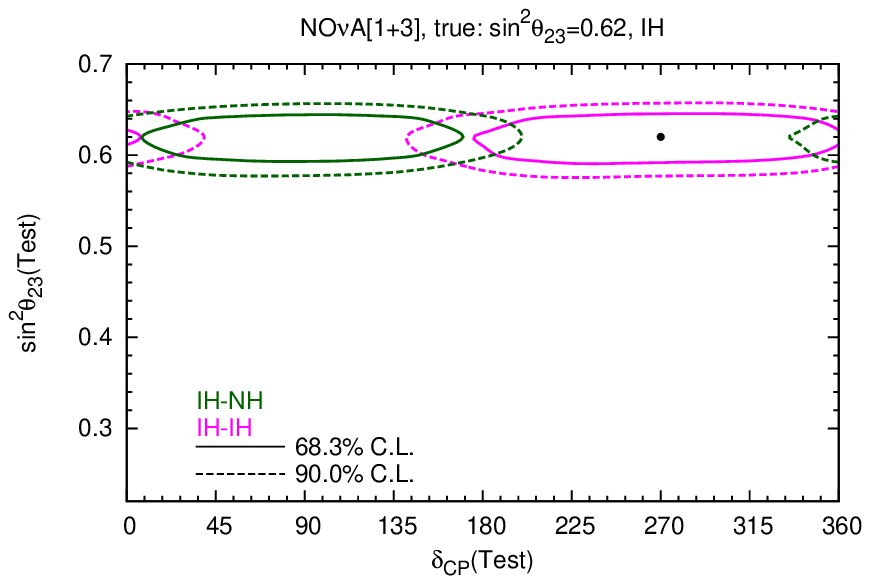} \\
     \hspace*{0.3in} 
  \includegraphics[height=4.5cm,width=5.5cm]{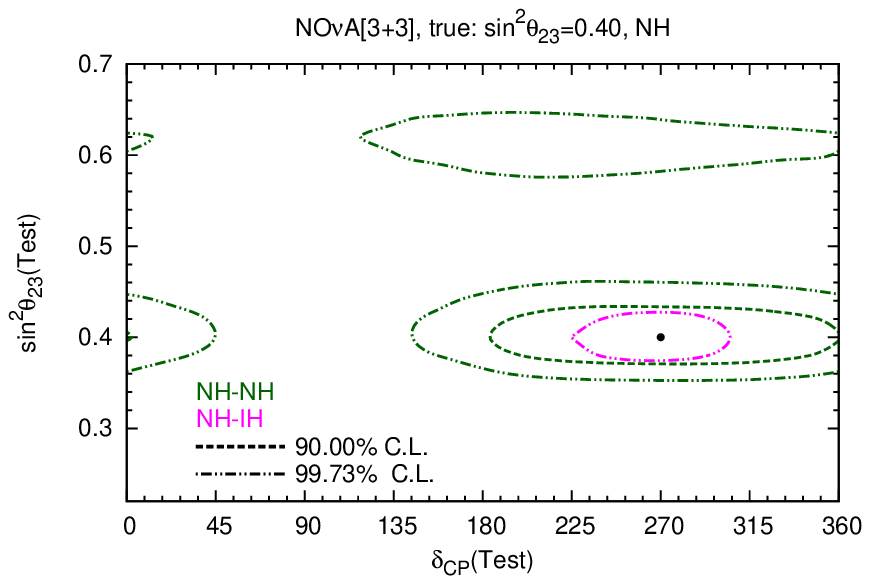} 
   \hspace*{-0.1in} 
 \includegraphics[height=4.5cm,width=5.5cm]{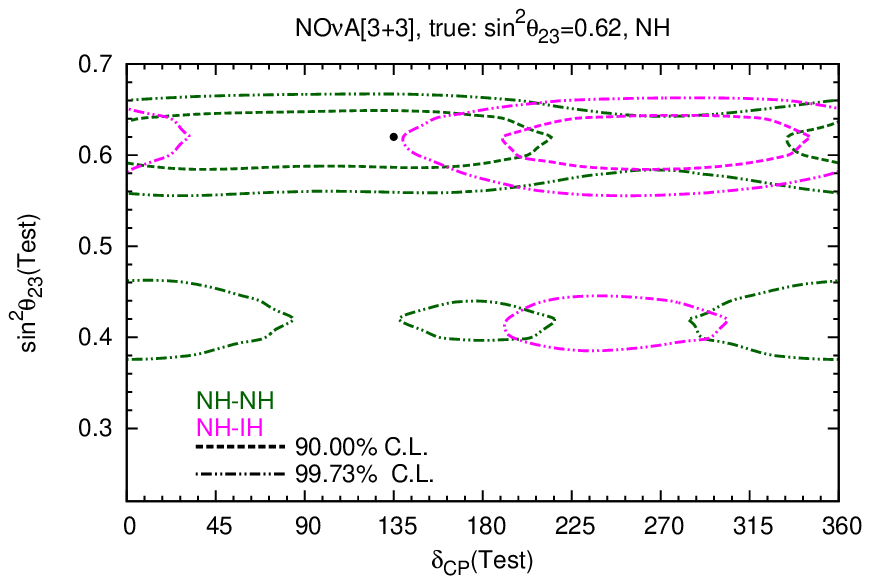} 
   \hspace*{-0.1in} 
  \includegraphics[height=4.5cm,width=5.5cm]{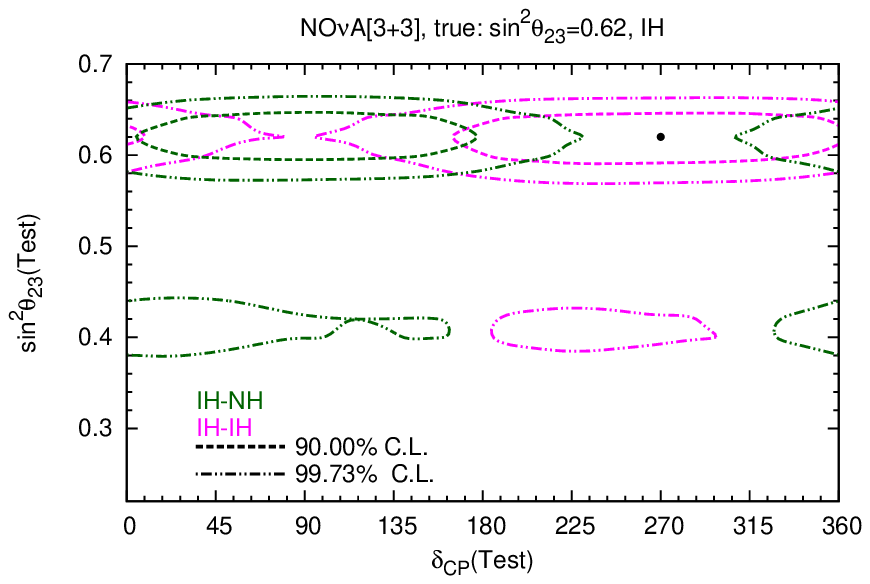}
 \end{tabular}
\caption{ Allowed contours in the test $\sin^2\theta_{23}-\dcp$ plane 
with the 
top, middle and bottom rows representing 3+1, 1+3 and 3+3 years of ($ \nu + \bar{\nu} $) runs respectively.
Here, for the first and second column green (magenta) curves are for  true-test 
hierarchies as NH-NH
(NH-IH) whereas for the third column magenta (green) curves are for IH-IH (IH-NH). Thus, the magenta(green) contours in the first two (third) 
column represent the WH solution.} 
\label{fig:nova_anu_1_3_3}
\end{figure*}

To asses this,  in the first row of figure \ref{fig:nova_anu_1_3_3}
we show the plots for 3+1 (i.e 3 years of neutrino and 1 year of antineutrino run) 
case whereas in the 2nd row we demonstrate the allowed 
area for the 1+3 (i.e 1 year of neutrinos and 3 years of antineutrino run) 
case.  In these figures (and all subsequent figures) we  also explore the
occurrence of the 
wrong hierarchy solutions by giving the contours for wrong hierarchy 
in the same plot unlike the figure~\ref{fig:nova_1_0_n_1_1}. 
The first plot in the top row is  for the true value in the LO and $\dcp = 270^\circ$. It shows that for 3+1 years of \nova run the WO solution gets largely
reduced at 90\% C.L. The precision of the true solution is also increased. 
For this case, no wrong hierarchy solution appears at 90\% C.L.  
However, if the runtime is 1+3 years then the wrong octant solution is 
resolved at 90\% C.L. though the precision around the true solution 
is better for the previous case due to more statistics. 

The middle column is for $\dcp \sim 135^\circ$ and HO with true hierarchy 
as NH. The first row (3+1 case) shows the presence of WH-RO-W$\dcp$ solutions
(magenta contours) even at 68\% C.L. A small region corresponding
to RH-WO solution is also seen to be 
present at 90\% C.L. The 2nd row shows the contours for the 1+3 case. 
In this case, a small RH-WO region as well as a small WH-WO region 
is seen to be present at 90\% C.L. 

The last column is for true hierarchy IH, $\dcp = 270^\circ$ and 
$\sin^2\theta_{23} =  0.62 $. In this case also, for 3+1 scenario the 
WH-RO-W$ \delta_{CP} $ solutions are visible even at 68\% C.L.    
A very  small region with wrong octant also remains allowed at 90\% C.L.       
For the 1+3 case (2nd row) the WO solutions are resolved 
at 90\% C.L.

The last row is for the 3+3 case which is the projected run plan of 
\nova. For this case we present the contours at 90\% and 99.73\% C.L. 
From the first plot in the last row we see that the WO-RH solution is resolved 
at 90\% C.L. but appears at 99.73\% C.L. 
In addition a WH-RO-R$ \delta_{CP} $ solution (magenta dash-dotted contour) 
is also seen at 99.73\% C.L. In the middle panel of the  last row the WH-W$ \delta_{CP} $-RO solutions can 
be seen at 90\% C.L. In addition, WH-WO-W$ \delta_{CP} $ as well as WO-RH solution is seen at 99.73\% C.L. 
Similar results can also be found for the third column for which true hierarchy 
is IH. 

Thus we can conclude that 1+3 case will do relatively better for removing 
the WO solutions at 90\% C.L. for all the three cases.  
However, for the solutions with true value in the HO, the WH-RO solutions 
remain unresolved.  Even with 3+3 runtime also, 
the WH-RO-W$ \delta_{CP} $ solutions 
are not resolved for  the two scenarios where the  true
value of  $\theta_{23}$ is considered to be in the higher octant. 

In view of this it is instructive to see to what extent the DUNE  experiment will be able to resolve the degeneracies. This is addressed in the next section. 

\begin{figure*}[htbp]
 \begin{tabular}{lr}
 \includegraphics[height=5cm,width=7cm]{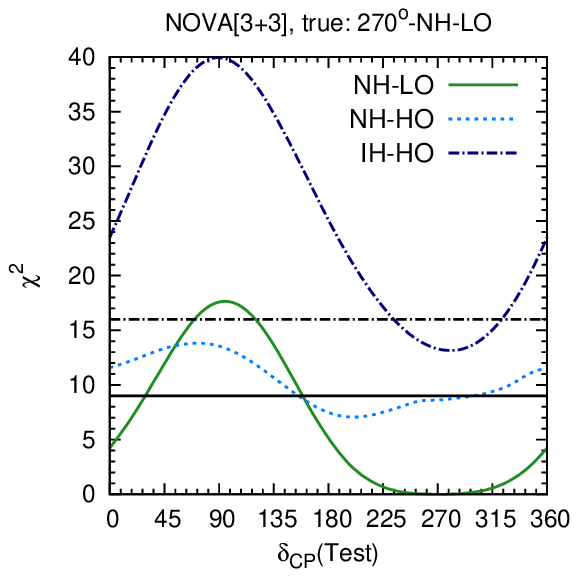} 
   \hspace*{-0.9in} 
  \includegraphics[height=5cm,width=7cm]{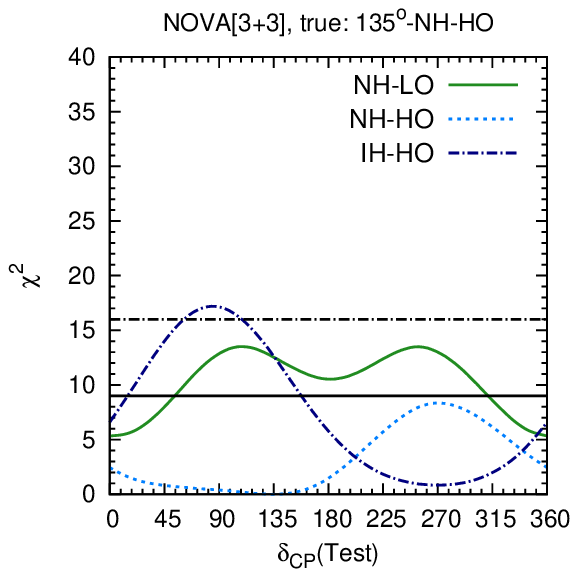} 
   \hspace*{-0.9in} 
 \includegraphics[height=5cm,width=7cm]{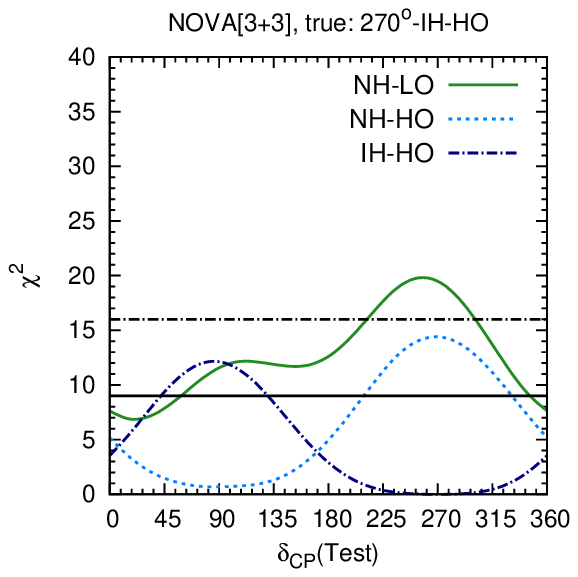} 
 \end{tabular}
\caption{CP sensitivity for NO$ \nu $A[3+3]. True points corresponding to left, middle and right columns are NH-LO, NH-HO and IH-HO respectively.}
\label{fig:cp_sen_nova}
\end{figure*}

In figure~(\ref{fig:cp_sen_nova}), we plot the CP sensitivity
for NO$ \nu $A[3+3]  as a function of 
test $\dcp$. The left most column is for true NH-LO. 
We notice  from this plot that that if the true solution is NH-LO then  
NO$ \nu $A[3+3] can rule out the NH-HO solution
at 95.45\% C.L. ($ \chi^{2} = 4 $) whereas  the range 
$156^\circ <\delta_{CP} < 315^\circ$ remains allowed at 99.73\% C.L. 
If this is the true solution then the degeneracies are more prominent for 
the right value of $\dcp$. 
It is also seen that the 
IH-HO solution  gets disfavoured 
at 99.73\% C.L. ($ \chi^{2} = 9 $). 
The middle panel is for true NH-HO. In this case the IH-HO 
solution gives a degenerate minima near  $\dcp \sim 270^o$ 
but 
is disfavoured at 99.73\% C.L. for most of the 
$\dcp$ values in the same half plane as the 
true $\dcp$. 
The NH-LO solution is disfavoured at 99.73\% C.L.  
excepting for $\dcp$ near the CP conserving values of $0$ and $\pi$. 
The third panel is for true IH-HO. In this case NH-HO 
gives degenerate minima with wrong $\dcp$ values.  
The NH-LO solution, on the other hand, is disfavoured at 99.73\% C.L. for 
most of the $\dcp$ values. 
Although the information content of
figure~(\ref{fig:cp_sen_nova}) is similar to the panel 3 of 
figure~\ref{fig:nova_anu_1_3_3}, the former shows precisely with what C.L. 
the wrong solutions are disfavoured for all values of  $\dcp$.

\section{Results  for NOVA+DUNE}\label{sec:nova_dune}
\begin{figure*}[htbp]
\vspace{-1.0cm}
 \begin{tabular}{lr}
 \hspace*{0.3in} 
 \includegraphics[height=4.5cm,width=5.5cm]{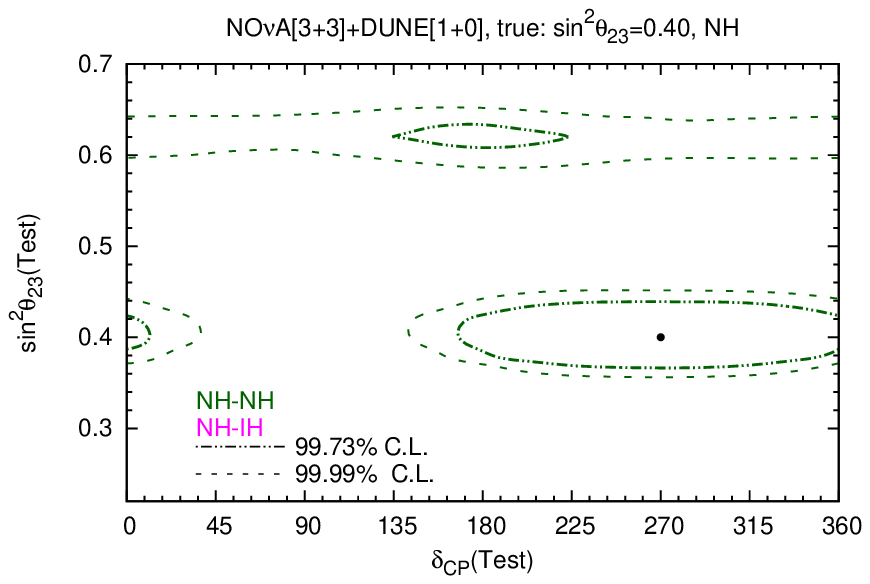} 
   \hspace*{-0.1in} 
  \includegraphics[height=4.5cm,width=5.5cm]{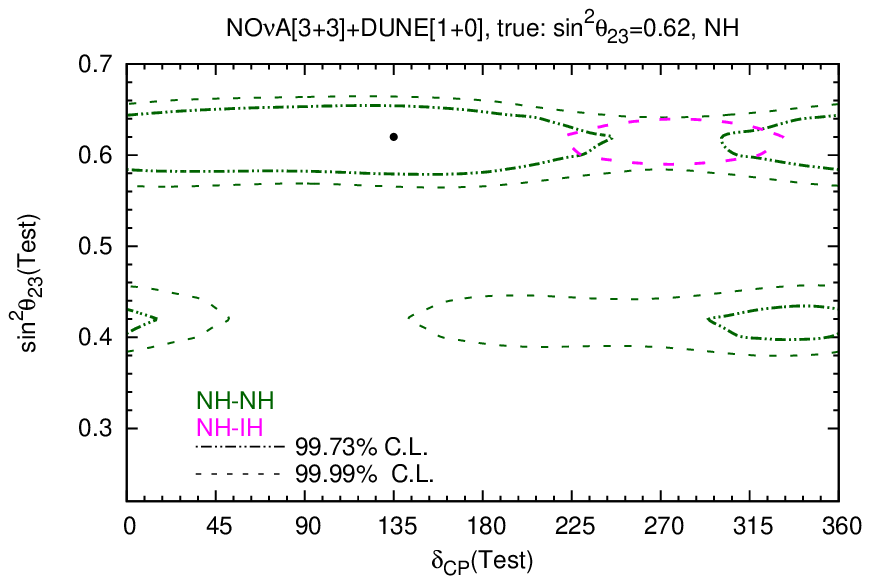} 
   \hspace*{-0.1in} 
 \includegraphics[height=4.5cm,width=5.5cm]{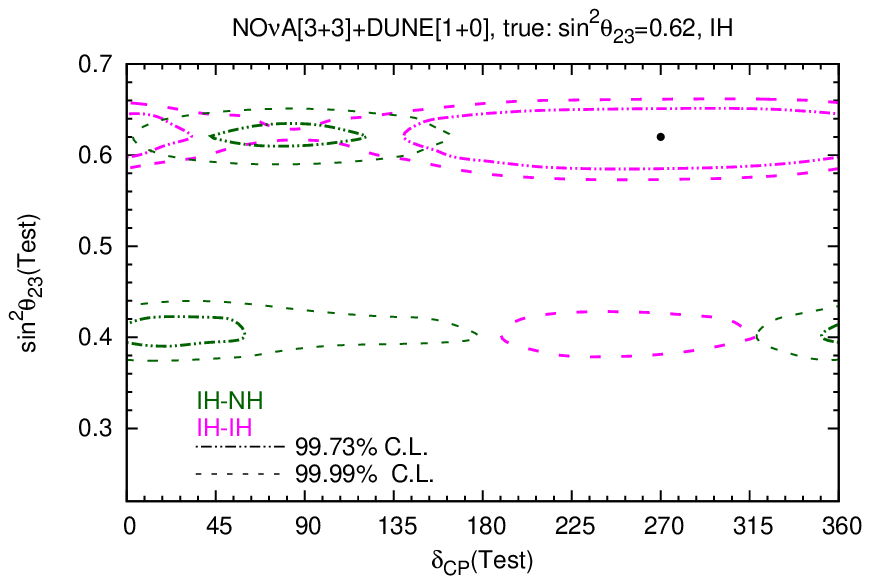} \\
  \hspace*{0.3in} 
  \includegraphics[height=4.5cm,width=5.5cm]{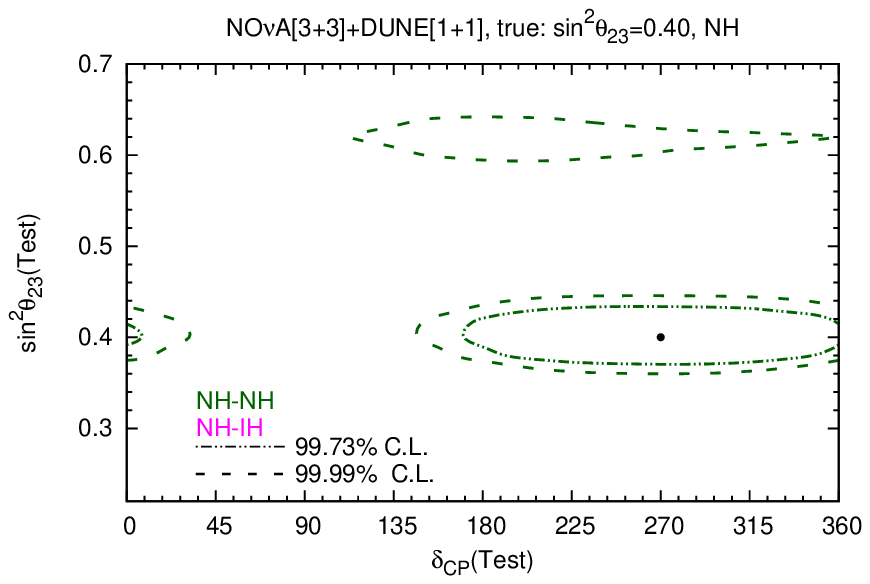} 
   \hspace*{-0.1in} 
  \includegraphics[height=4.5cm,width=5.5cm]{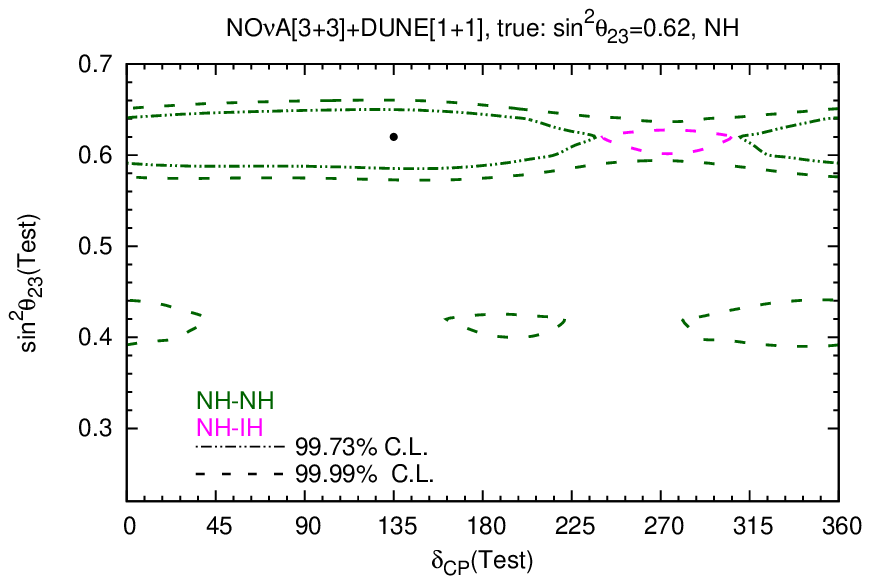} 
   \hspace*{-0.1in} 
 \includegraphics[height=4.5cm,width=5.5cm]{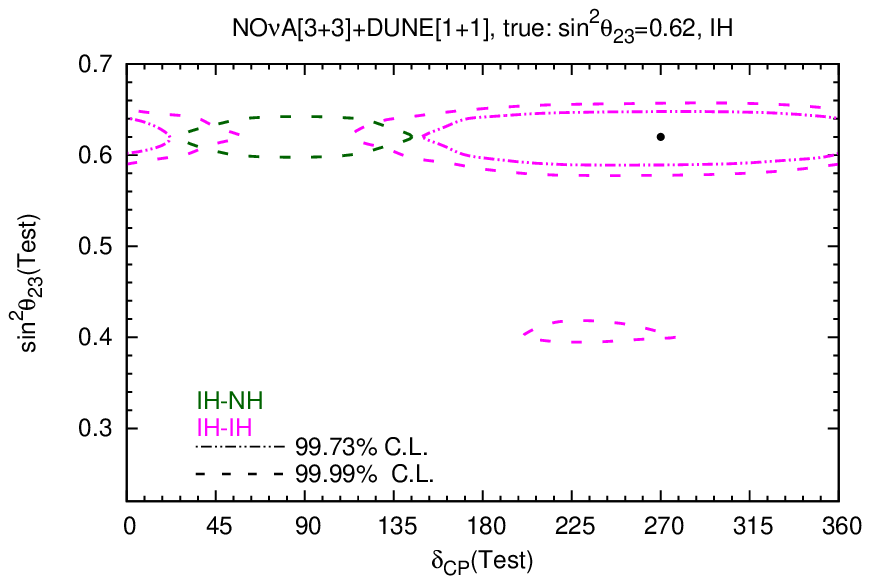} \\
  \hspace*{0.3in} 
   \includegraphics[height=4.5cm,width=5.5cm]{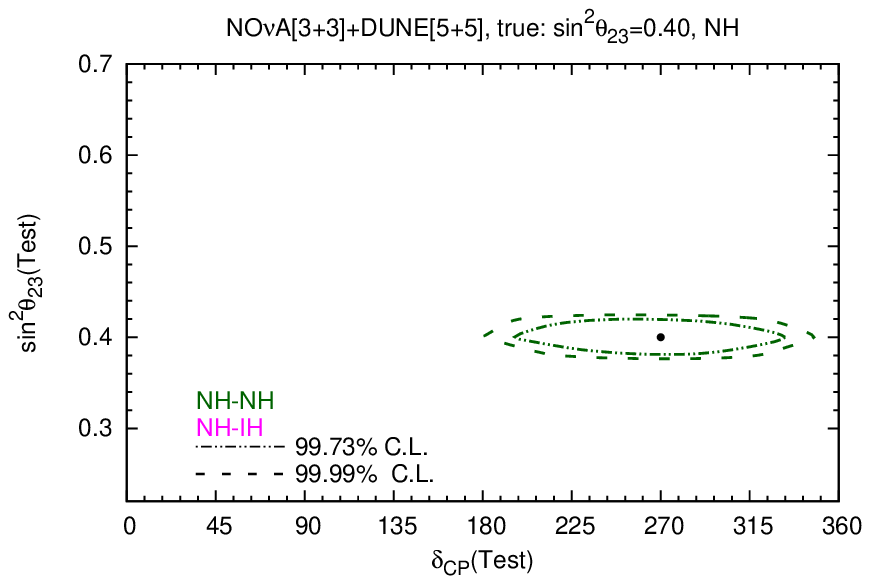} 
   \hspace*{-0.1in} 
  \includegraphics[height=4.5cm,width=5.5cm]{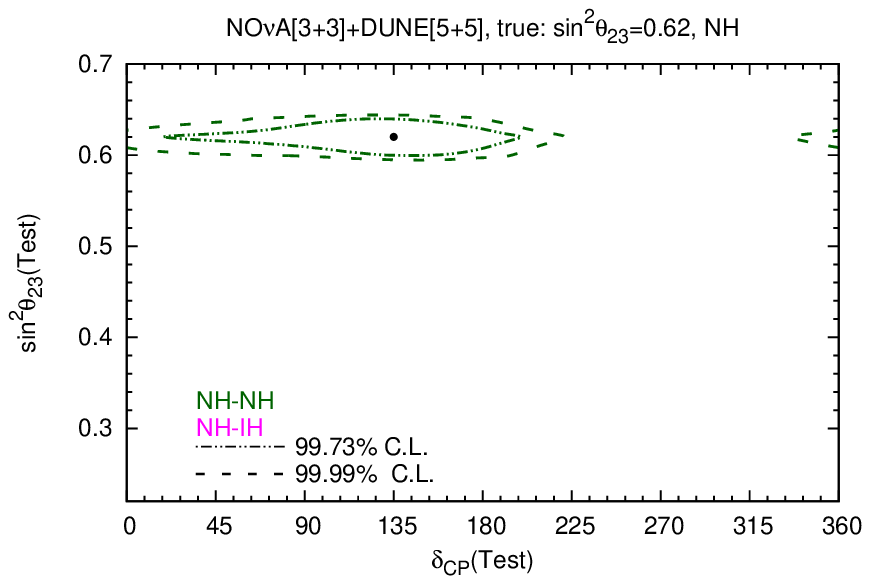} 
   \hspace*{-0.1in} 
 \includegraphics[height=4.5cm,width=5.5cm]{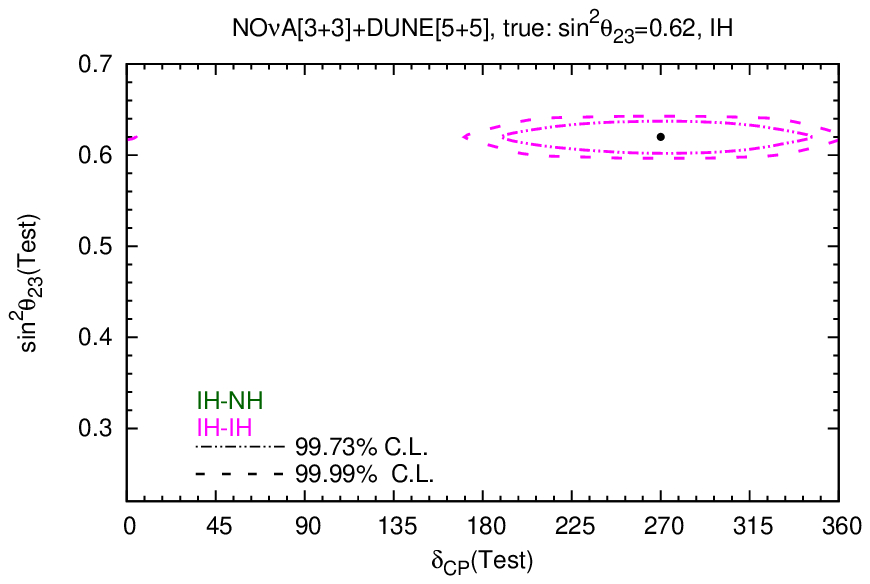}
 \end{tabular}
\caption{Same as figure \ref{fig:nova_anu_1_3_3} but top, middle and bottom row are for 1+0, 1+1 and 5+5 years of ($ \nu + \bar{\nu} $) runs of DUNE  respectively, combined with \nova[3+3].}
\label{fig:nova_dune}
\end{figure*}

In this section we discuss to what extent the DUNE experiment can resolve 
the degeneracies existing after \nova 3+3 run. 
The first row of figure~\ref{fig:nova_dune} demonstrates the results after 1 year of neutrino run of DUNE while the second row is for 1+1 years of ($ \nu + \bar{\nu} $) run.
The last row presents the results
for  5+5 years   ($ \nu + \bar{\nu} $) run of DUNE. 
DUNE, being a high statistics experiment,
we present the contours at 
99.73\% ($3\sigma$) and 99.99\% (4$\sigma$)  C.L. for all  the panels.  

The first plot shows the presence of RH-WO solution at 99.99\% C.L. 
with no wrong hierarchy solution as expected. The full range of 
$\dcp$ remains allowed at this C.L.  There is also a small 
WO-RH solution at 99.73\% C.L. The 1st plot in the middle row
shows the same results with 1+1 year run of DUNE. For this 
case the WO-RH solution appears at 99.99\% C.L. around right $\dcp$. 

The middle column in this plot is for $\dcp =135^\circ$ and HO.
In this case with the addition of 1 year of neutrino data from DUNE 
to \nova[3+3], the WH-RO-W$ \delta_{CP} $ 
solution is seen to be resolved at 99.99\% C.L. However, 
a small WO solution remains at this C.L. around $\dcp=360^\circ$. 
The middle plot in the 2nd row shows that with 1+1 run of DUNE 
there are no WO and WH solutions at 99.73\% C.L. 
However,  WH-RO-W$ \delta_{CP} $ and RH-WO solutions appear at
99.99\% C.L.
 
The last column represents true IH, HO and $\dcp = 270^\circ$. 
For this case WH-RO and RH-WO solutions are seen at 99.73\% C.L. 
in the first row i.e for 1+0 run of DUNE. 
But with 1+1 these solutions appear at 99.99\% C.L. only. 

The last row is for DUNE (5+5).  One can see that for this 
case 
all the degenerate solutions are resolved by DUNE at 99.99\% C.L. . 

\begin{figure*}[htbp]
        \begin{tabular}{lr}
 \hspace*{0.3in} 
\includegraphics[height=4.5cm,width=5.5cm]{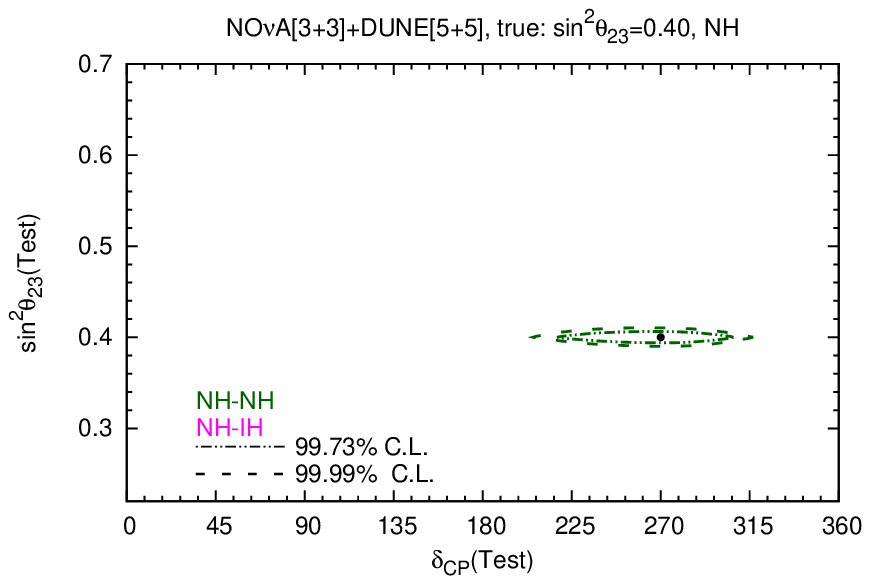}
   \hspace*{-0.1in}
\includegraphics[height=4.5cm,width=5.5cm]{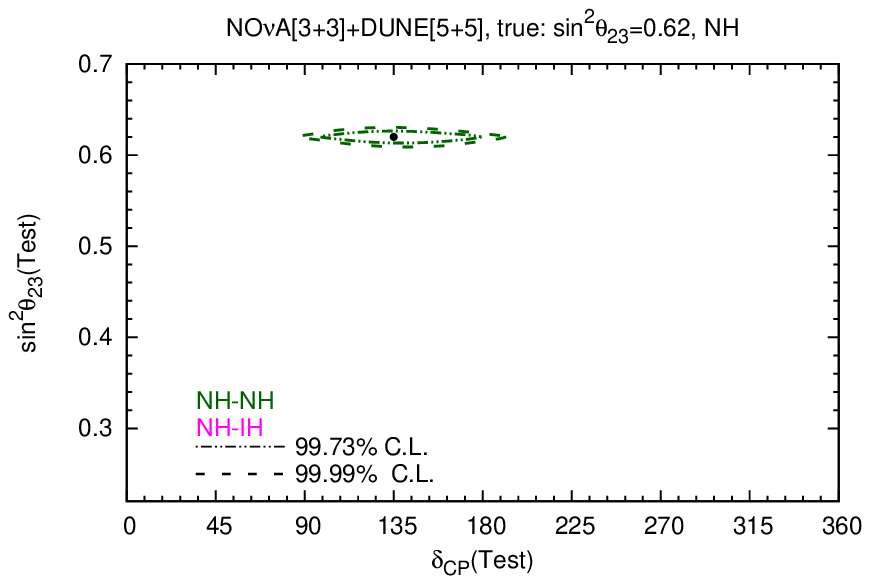}
   \hspace*{-0.1in}
\includegraphics[height=4.5cm,width=5.5cm]{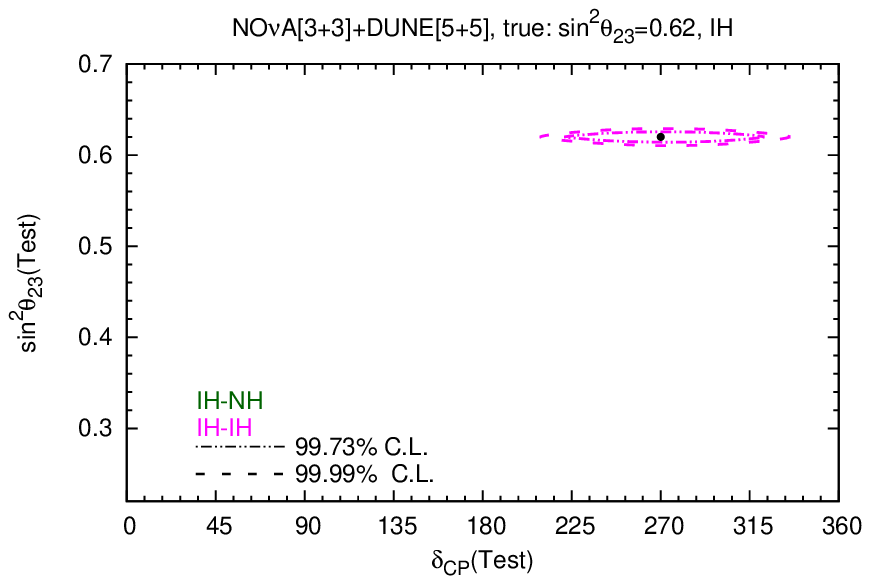}
        \end{tabular}
\vspace{-0.35cm}
\caption{The allowed areas in test $\dcp - \sin^2\theta_{23}$ plane 
for  NOVA[3+3]+DUNE[5+5], using DUNE detector volume as 40 kton.}
\label{fig:nova_dune_40kt}
\end{figure*}

In figure~\ref{fig:nova_dune_40kt} we show the  allowed areas in the 
test $\dcp - \sin^2\theta_{23}$ plane for the second phase of DUNE with a volume of 40 kton. 
We can see that the precision improves considerably with enhanced statistics. 
Specially the demarcation between the CP conserving and CP violating values 
are seen to be much improved. The precision of $\theta_{23}$ also shows 
significant improvement. In table~(\ref{table:precision}) we 
compare  the  precision of   $\sin^{2}\theta_{23} $ and $ \delta_{CP}$,
around the best-fit values, for the two cases.  
The precision of these two parameters can be defined as: 
\begin{align}
{\rm Precision} &=  \dfrac{Max - Min}{{\rm(true~value)} \times 6} \times 100\%
\end{align}
It is seen that below 1\% precision in $\sin^2\theta_{23}$ and 
$\lsim 10\%$ precision in $\dcp$ can be achieved by DUNE 40 kton 
volume for the true values considered. 
\begin{table}
\begin{tabular}{| c | c | c | }
\hline
   & DUNE 10 kton & DUNE 40 kton \\
   &  (in \%) &   (in \%) \\
\hline
True pt. &  $\sin^{2}\theta_{23}~~~~~~ \delta_{CP} $ & $\sin^{2}\theta_{23}~~~~~~\delta_{CP} $ \\
\hline
0.40-270-NH & 1.6 $ ~~~~~~~~~~$  9 & 0.6$~~~~~~~~~~~$ 6  \\
0.62-135-NH & 1 $ ~~~~~~~~~~$  22 & $~~$ 0.5 $~~~~~~~~~~$ 11 \\
0.62-270-IH & 0.8 $~~~~~~~~$ 10  &$~~$ 0.4 $~~~~~~~~~~$ 6\\
 \hline
\end{tabular}
\caption {Precision table on $\sin^{2}\theta_{23}$  and $ \delta_{CP} $ for NO$ \nu $A[3+3]+DUNE[5+5] with 10 kton and 40 kton detector mass of DUNE respectively.}
\label{table:precision}
\end{table}

\section{Conclusions} \label{sec:conclusion}
We discuss the implications of the recent \nova results 
for future runs of \nova as well as for DUNE. 
The best-fit parameters reported in \cite{Adamson:2017gxd} 
point towards degenerate solutions.  
We  explore  to what extent these degeneracies can be  resolved
by tuning the neutrino and antineutrino  run  times of NO$ \nu $A.
We consider the latest \nova best-fit values as our true values.  
First we describe the effect of 1 year of antineutrino run of \nova
and find that the wrong octant solutions can be resolved at 68\% C.L. 
for all the three choices of true values. 
Next we present the comparison of \nova 1+3 ($\nu+\bar{\nu}$) and
3+1 case 
and find that 1+3 provides a better option for the removal of 
wrong-octant solutions. However, the wrong hierarchy-wrong $\dcp$ 
solution remains present at 90\% C.L.  
even after 3+3 year running of \nova.   
DUNE[1+1] can resolve the degeneracies at 99.73\% C.L. 
whereas DUNE[5+5] can resolve all the  degeneracies at 99.99\% C.L. 
In conclusion, if one of the current \nova best-fit values in  reference 
\cite{Adamson:2017gxd} is the 
true value, then the FERMILAB experiments \nova and DUNE together will be 
able to solve all the  degeneracies at 99.73\% C.L. 
The second phase 
of DUNE with a volume of 40 kton can further improve the precision of the 
allowed regions.     

\section{Acknowledgement} 

{S.G. wants to thank Monojit Ghosh for many helpful discussions}. 

\bibliography{neutosc}
\end{document}